# MARS IN THE AUSTRALIAN PRESS, 1875–1899. 2. CIRCULATION AND ATTRIBUTION

**Richard de Grijs**
School of Mathematical and Physical Sciences, Macquarie University,
Balaclava Road, Sydney, NSW 2109, Australia
Email: richard.de-grijs@mq.edu.au

**Abstract:** Between 1875 and 1899, Mars occupied a prominent and recurring position in newspaper reporting across Europe, North America and beyond. Although the scientific and cultural dimensions of this "Mars excitement" have been well studied in metropolitan contexts, far less attention has been paid to how planetary astronomy circulated through colonial press systems. This paper examines the Australian press as a case study in the global mediation of nineteenth-century astronomical knowledge. Drawing on a structured compilation of 1040 digitised newspaper articles accessed via the National Library of Australia, the study analyses patterns of first appearance, reprinting, attribution and temporal persistence in Australian Mars reporting. Mars-related news entered Australia primarily through international telegraphic networks and overseas syndication before circulating widely through metropolitan, regional and provincial newspapers. Distinguishing between novel reports and subsequent reprints reveals that apparent abundance often masked a small number of originating items that achieved extensive colonial reach. Attribution practices evolved over time, shifting from anonymous or institutional authority towards increasing reliance on named overseas figures such as Asaph Hall, William H. Pickering and, most notably, Percival Lowell. This shift reshaped both the form and longevity of Mars reporting, allowing interpretive and personality-driven material to persist independently of specific observational events. By emphasising circulation, attribution and temporal rhythm, this study situates Australian newspapers within international systems of scientific communication. Colonial journalism actively mediated astronomical knowledge rather than merely transmitting it, shaping how Mars became a durable object of public scientific attention in the late nineteenth century.

## 1. INTRODUCTION

Mars occupied a uniquely prominent position in late nineteenth-century astronomical reporting. From the great opposition of 1877 through the end of the century, the planet generated recurring waves of attention in newspapers across Europe, North America and beyond. Whereas the scientific and cultural dimensions of this "Mars excitement" have been extensively examined for metropolitan centres (e.g., Crowe, 1986; Sheehan, 1996; Crossley, 2011; Lane, 2011), its circulation and mediation within colonial press environments remain comparatively underexplored.

Here, I examine how Mars-related astronomy entered, circulated within and persisted across the Australian press between 1875 and 1899. Rather than focusing on the interpretive meanings associated with Mars—a subject addressed in Paper 1—this paper investigates the mechanisms of transmission that shaped public access to planetary astronomy. It asks how Mars news travelled to Australia, how it was reproduced and reframed and how scientific authority was constructed and maintained in print. Australian newspapers provide an especially revealing case study. By the late nineteenth century, they were fully integrated into global telegraphic networks while also serving geographically dispersed colonial readerships, receiving overseas scientific and general news via imperial cable systems and commercial news agencies (e.g., Moyal, 1984; Potter, 2012). This combination produced a press environment in which overseas scientific news could arrive rapidly, circulate widely and be selectively adapted to local contexts. Mars, with its recurring oppositions and interpretive openness, proved particularly well suited to this system.

Using a structured compilation of digitised newspaper items drawn from *Trove* at the National Library of Australia (see Section 2), this study traces patterns of first appearance, reprinting, attribution and temporal rhythm in Australian Mars reporting. Particular attention is paid to the shifting roles of named overseas scientific authorities—most notably Asaph Hall (1829–1907), William H. Pickering (1858–1938) and Percival Lowell (1855–1916)—and to the ways in which telegraphy and syndication shaped both the form and longevity of news.

By emphasising circulation, attribution and timing, this paper contributes to the history of astronomy by demonstrating how planetary (solar system) knowledge was mediated through colonial journalism. In doing so, it complements existing scholarship on Mars by situating astronomical excitement not only in observatories and books, but in the everyday mechanics of the press.

## 2. SOURCES, DATABASE CONSTRUCTION AND SEARCH STRATEGY

### 2.1 Trove and the Australian newspaper record

This study is based on digitised Australian newspapers accessed through *Trove*, the online discovery service maintained by the National Library of Australia (https://trove.nla.gov.au/). *Trove* provides full-text access to a large collection of nineteenth-century newspapers published across the Australian colonies, with coverage varying by title, region and period. For the late nineteenth century, *Trove* offers an unparalleled resource for tracing the circulation of scientific news and commentary within the colonial press, including metropolitan, regional and provincial newspapers.

Digitised newspapers are particularly well-suited to longitudinal studies of public science, as they allow the reconstruction of patterns of reporting over extended periods and across multiple publications. At the same time, the limitations of such sources are well-known, notably the variable quality of optical character recognition (OCR) applied to historical typefaces and degraded originals. These limitations do not invalidate the compilation's depth or quality but require that search strategies be designed with care and transparency. The temporal scope of this study extends from 1875 to 1899. The starting date precedes the great opposition of Mars in 1877 and allows for the identification of anticipatory or contextual material, while the endpoint captures the consolidation of Mars-related discourse in the Australian press before the major technological and cultural shifts of the early twentieth century.

The database consists of newspaper items relating to Mars that were identified through a structured keyword search strategy (see below) and then organised chronologically according to their first appearance in the Australian press. Where items were subsequently reprinted, syndicated or reproduced verbatim in other newspapers, these later appearances were recorded alongside the initial publication. In the master database file, newly appearing items are marked with an asterisk, followed by one or more newspaper references documenting subsequent reprints. This structure makes it possible to distinguish between novelty and repetition, and to examine how particular items propagated through the colonial press network over time. The raw data underlying this series of papers is available at https://astro-expat.info/Data/mars19c.html (for long-term preservation, see also https://web.archive.org/web/20260112080309/https://astro-expat.info/Data/mars19c.html).

### 2.2 Search strategy and OCR considerations

The construction of the article database relied on three principal search strategies applied within *Trove*:
- planet Mars canals
- Mars Schiaparelli
- "planet Mars" AND (Hall OR Pickering OR Lowell)

These searches were designed to capture complementary dimensions of Mars reporting: concept-driven discussion (canals), named scientific authority (Schiaparelli, Hall, Pickering, Lowell) and general astronomical framing ("planet Mars"). An important consideration in the formulation of these search terms was the practical limitation imposed by OCR accuracy in nineteenth-century newspapers. Searches based solely on the word "Mars" return an extremely large number of false positives owing to OCR confusion with typographically similar words such as "Mar", "Mare" and related letter strings. Such results rapidly become unmanageable and obscure relevant material. The deliberate inclusion of the term "planet" in two of the three search strategies served both a semantic and a practical function. Semantically, it restricts results to astronomical contexts; pragmatically, it significantly improves precision and reproducibility by filtering out most OCR-induced false matches. This approach represents a compromise between completeness and reliability that is appropriate for a structured historical analysis. Each of the three search strategies plays a distinct role in defining the article collection.

The search string "planet Mars canals" captures reporting in which the canal concept is explicitly brought into focus. Such items are often speculative or interpretive in tone and are particularly useful for tracing the public life of the canal hypothesis beyond specialist astronomical circles. The search string "Mars Schiaparelli" anchors reporting to the origin of the canal terminology and identifies items that situate Mars observations within an explicitly European scientific lineage. These results are especially valuable for tracking how Schiaparelli's work was described, translated or reinterpreted in the Australian press. The compound search |"planet Mars" AND (Hall OR Pickering OR Lowell)| captures reporting associated with named individuals whose authority derived from observational discovery (Hall), organised observing programmes (Pickering) or interpretive synthesis and popularisation (Lowell). Together, these names span the period from the initial surge of interest following the favourable 1877 opposition to the interpretive debates of the 1890s (see Paper 1).

Although the primary search strategy focused on terms directly associated with observational claims about Mars, prominent popularisers such as Camille Flammarion (1842–1925) also appeared frequently in the discourse. References to Flammarion tended to frame Mars within broader speculative and philosophical discussions of extraterrestrial life rather than within the narrower observational debates that structure this analysis. For this reason, such material was incorporated contextually where relevant but was not used as a primary organising category.

Whereas no keyword strategy can capture every relevant reference to Mars, the combination of these searches produces a compilation that is both manageable and representative of the principal modes through which Mars entered Australian newspaper discourse. Items retrieved through these searches were visually inspected to confirm their relevance to Mars as an astronomical object. Purely metaphorical, poetic or incidental references to "Mars" unconnected to astronomy were excluded, as were items in which Mars appeared only as part of a generic list of planets (e.g., in the context of monthly stargazing columns) without substantive discussion. The resulting database, containing 1040 articles of relevance, does not claim exhaustive coverage of every mention of Mars in the Australian press between 1875 and 1899. Rather, it provides a structured and transparent sample that is well-suited to analysing patterns of transmission, attribution and authority. The strengths and limitations of this approach will be discussed further in Section 8.

This paper focuses on the mechanisms by which Mars-related astronomical information circulated through the Australian press. The interpretive consequences of that circulation—including public excitement, speculation and local resonance—are examined in Paper 1, which draws on the same article database but addresses a different set of historical questions.

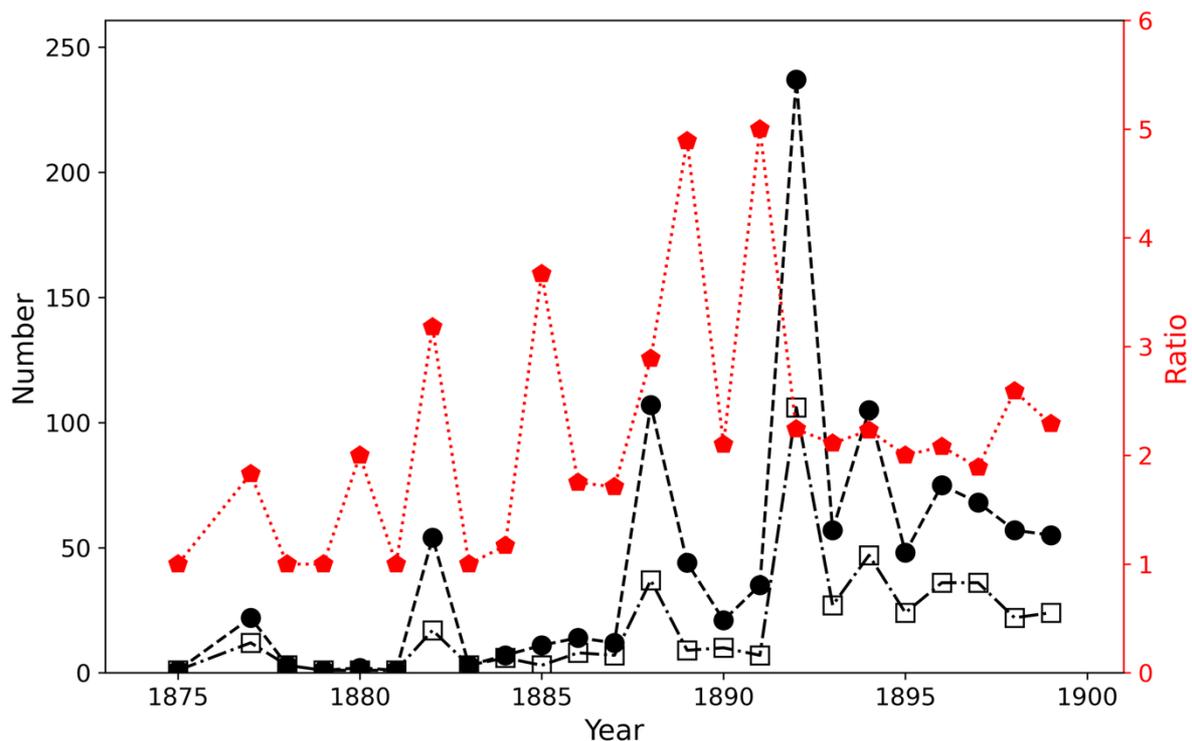

**Figure 1.** Visual representation of the temporal dependence of Australian Mars reporting. The solid black bullets connected by dashed lines show the number of articles identified as a function of year of publication (first appearance); the open squares connected by dash-dotted lines represent the number of article clusters. Their ratios, reflected by the quantitative scale on the right-hand *y* axis, are shown by the red pentagrams connected by dotted lines. The favourable oppositions of 1877, 1882, 1888, 1892 and 1894 are clearly identifiable.

## 3. FIRST APPEARANCE, REPETITION AND SYNDICATION

### 3.1 Identifying novelty in the colonial press

A central aim of this study is to distinguish between novel reporting and subsequent repetition/reprinting within the Australian press. Nineteenth-century newspapers routinely reprinted material from other colonial, British and American sources, often without attribution or with minimal editorial intervention

(e.g., Cryle, 1997; Secord, 2004; Potter, 2012). As a result, the apparent abundance of Mars-related items can be misleading unless first appearances are clearly identified.

In the present compilation, items are therefore organised chronologically by their earliest identifiable publication date in the Australian press. As established in Section 2, distinguishing first appearances from subsequent reprints is essential to reconstruct the life cycle of certain news items, from first arrival to eventual saturation and decline. An early example of such a first appearance occurs in the *Ipswich Observer and West Moreton Advocate* of 24 December 1875, which contains one of the earliest Mars-related items in the database. Although modest in scope, this item establishes a baseline against which the dramatic increase in reporting following the 1877 opposition can be measured. Mars excitement in the Australian press followed the natural rhythm of major and minor oppositions. Significant increases in press coverage occurred in 1877, 1882, 1888, 1892 and 1894, whereas the baseline level gradually increased over the period of interest

## 3.2 Patterns of initial appearance, reprinting intensity and temporal persistence

Analysis of first appearances reveals that Mars-related news entered the Australian press in episodic bursts rather than as a steady stream: see Figure 1. These bursts correspond closely to identifiable external triggers, most notably favourable oppositions of Mars. A contrasting pattern is visible in the late 1890s, when Mars-related items increasingly enter the Australian press not in response to discrete observational events but through interpretive commentary associated with named individuals. For example, items reporting Lowell's views on Martian canals, first appearing in Australian newspapers in the mid-to-late 1890s, often arrive as extended essays or summaries of lectures rather than brief observational notices (e.g., *The Australasian*, 12 October 1895; *Argus*, 16 November 1895; *Maitland Daily Mercury*, 4 February 1896; and numerous reprints). These items are less tightly clustered around oppositions and instead exhibit a slower but more sustained pattern of circulation. This shift reflects a broader change in the nature of Mars reporting, from event-driven astronomy to personality-driven interpretation.

The impact of the 1877 opposition is immediately apparent in the article collection. One of the earliest Australian reports appears in the *Sydney Morning Herald* of 11 August 1877, closely followed by reports in regional and interstate newspapers such as the *Dalby Herald and Western Queensland Advertiser* (8 September 1877) and the *Argus* (Melbourne; 16 October 1877). The clustering of first appearances within a narrow temporal window reflects the rapid uptake of overseas astronomical news during periods of heightened interest. In several cases, multiple first appearances occur on the same day in different newspapers, suggesting near-simultaneous access to cable reports rather than a simple linear chain of reprinting or syndication within Australia.

Once introduced, Mars-related items often exhibit extended persistence in the press. Items originating in metropolitan newspapers are frequently reprinted in regional and provincial titles, sometimes verbatim and sometimes in abridged form. For example, reports first appearing in the *Argus* in October 1877 were subsequently reproduced in other colonial newspapers over the following weeks, often without explicit acknowledgement of their origin. Such repetition creates the impression of widespread and sustained novelty, even when the underlying information derives from a single source. This persistence is especially pronounced for items that combine observational reporting with speculative or explanatory content. By contrast, narrowly technical notices tend to attract fewer reprints and disappear more rapidly from the press record. A clear example of this later pattern can be seen in reports summarising Percival Lowell's claims regarding the artificial nature of the Martian canals. Rather than presenting fresh observational data, these items typically offered a compact synopsis of Lowell's interpretation: that the canals formed a geometrically precise, planet-wide network, incompatible with natural geological processes and best explained as vast irrigation works constructed by an intelligent but water-stressed Martian civilisation.

One such item, first published in an Australian metropolitan newspaper in the mid-1890s (*The Australasian*, 12 October 1895), exemplifies this mode of reporting. The notice reiterated Lowell's core arguments—emphasising the canals' straightness, global scale and apparent seasonal darkening—while framing them as the considered conclusions of a well-equipped and authoritative observer rather than as tentative or contested hypotheses. Crucially, the article did not anchor these claims to a specific observing night or planetary opposition but instead presented them as a settled interpretive claim about Mars itself.

Subsequently reprinted across multiple Australian colonies over an extended period, this item circulated in a markedly different temporal context from earlier, observation-driven reports. It continued to reappear weeks or even months later in regional newspapers, often detached from any immediate astronomical event. This persistence suggests that, by this stage, Mars-related reporting had gained a degree of discursive autonomy. The canals functioned less as a transient scientific finding than as a durable explanatory narrative—one that could be redeployed whenever editorial space permitted, regardless of whether new astronomical data were available. In this sense, Lowell's interpretation helped transform Mars from an episodic object of observation into a standing topic of speculation, debate and imaginative engagement within the colonial press.

### 3.3 Modification, condensation and reframing

Although many reprinted items appear verbatim, others undergo condensation or reframing as they circulate. Headlines are frequently altered to emphasise discovery or significance, and longer articles may be shortened to fit available space. In some instances, explanatory or speculative passages are retained while observational detail is reduced, thus subtly shifting the balance of the original report. Such editorial interventions are rarely signalled to readers, but their cumulative effect is to privilege interpretive narratives over methodological nuance. The ability to group reprints with their first appearances makes it possible to detect these changes and to assess how Mars-related astronomy was adapted for different audiences.

Late-1890s reprints also show a marked tendency towards reframing and headline amplification. Reports of Lowell's views are frequently introduced under headings that highlight speculation or controversy—such as the habitability of Mars or the intelligence of its putative inhabitants—while the observational basis of these claims is summarised only briefly or omitted altogether. A report derived from Lowell's writings that appeared in *The Australasian* (12 October 1895), for example, was subsequently reprinted with headlines that shifted emphasis away from telescopic observation and towards interpretive speculation. In regional newspapers, the same core text circulated under titles that invited readers to engage with questions of extraterrestrial life rather than with the technical grounds on which Lowell's claims rested. For example, substantially identical text to that published in *The Australasian* (Melbourne; 12 October 1895) appeared under the headline *"Is Mars Inhabited?"* or *"The Planet Mars. Is It Inhabited?"* in the *Evening Observer* (Brisbane; 12 October 1895), the *Maryborough Chronicle, Wide Bay and Burnett Advertiser* (Queensland; 17 October 1895), the *Glen Innes Examiner and General Advertiser* (New South Wales; 18 October 1895) and the *Richmond River Herald and Northern Districts Advertiser* (New South Wales; 25 October 1895). In other outlets the same material was presented as *"The Planet Mars"* (Western Australia; *Geraldton Murchison Telegraph*, 1 November 1895) or *"Mars and Its Inhabitants"* (Victoria; *Port Fairy Gazette*, 18 October 1895). In all cases, the body text closely followed the metropolitan original, but the headlines redirected reader attention from Lowell's observational claims to the broader question of intelligent life beyond Earth.

In some instances, closely related summaries of Lowell's canal observations appeared under markedly different forms of editorial framing, indicating active local mediation rather than uniform mechanical reprinting. At one end of the spectrum, Lowell's claims about the canals' straightness and seasonal darkening could be presented as restrained astronomical commentary. An example is *"The Planet Mars. Some Recent Discoveries"* (Launceston *Daily Telegraph*, 4 January 1895), which reports recent observations in a largely technical context, emphasising telescopic conditions, comparison with earlier observers and the limits of inference. Although the possibility of artificial structures is noted, it is not highlighted, and the article conforms closely to the conventions of routine science reporting.

Elsewhere, similar Lowell-derived material was embedded within far more speculative narrative frameworks. In *"Our Neighbours in Mars"* (*South Australian Chronicle*, 20 April 1895), the canals are again described in terms of their scale, regularity and apparent seasonal behaviour, but these observations are explicitly interpreted as evidence of a civilisation responding to environmental decline on an ageing planet. Mars is characterised as an "old" or desiccating world, and the canals are framed as deliberate engineering works necessitated by dwindling water resources. Here, the observational content is compressed, while its interpretive implications are amplified.

Taken together, these examples suggest that Mars-related reporting in the mid-1890s was no longer governed solely by the conventions of observational astronomy. Even where the same core claims about canals circulated, their significance could be reshaped through headline choice, narrative emphasis and contextual framing. Rather than functioning simply as transient reports tied to specific observing conditions, Lowell's canals became a flexible explanatory motif, capable of supporting sober scientific discussion or broader speculative reflections on planetary history and civilisation.

This editorial flexibility suggests that Mars reporting had become detached from the conventions governing routine astronomical notices. Rather than functioning solely as science reporting tied to specific observations or events, these items occupied an intermediate space between scientific commentary and speculative feature writing, reinforcing the emergence of Mars as a semi-autonomous and enduring topic within the late nineteenth-century press. This pattern contrasts sharply with the relatively restrained treatment of earlier Mars reports and illustrates how reprinting practices evolved as the subject moved from observational astronomy into the realm of public debate. Reprinting patterns also illuminate the geographic reach of Mars-related news within Australia. Items first published in major metropolitan centres such as Sydney or Melbourne frequently appear soon afterwards in regional newspapers in New South Wales and Victoria, but also in Queensland, South Australia, Western Australia, Tasmania and elsewhere, reinforcing the sense that Mars was a topic of shared colonial interest. The recurrence of the same items across geographically dispersed newspapers suggests that Australian readers encountered Mars not as an isolated curiosity, but as part of a sustained and widely distributed discourse.

## 4. ATTRIBUTION, ANONYMITY AND AUTHORITY

### 4.1 Attribution practices in nineteenth-century astronomical reporting

Attribution practices in nineteenth-century Australian newspapers were highly variable and often opaque. Astronomical items were frequently introduced under generic headings such as "Astronomical Notes", "Science Intelligence" or "Scientific Items", with little indication of their original source beyond vague references to overseas authority. Reports might be attributed simply to "a well-known observer" (e.g., *Mercury*, 23 August 1892) or a "recent investigation" (e.g., *Bendigo Advertiser*, 10 October 1896; *Prahran Telegraph*, 24 July 1897; and subsequent reprints), without naming either an individual or an institution. In other cases, material was said to derive from "scientific journals" (e.g., *The Age*, 24 September 1892; *Bathurst Free Press and Mining Journal*, 6 February 1893; *Argus*, 22 September 1894), "a contemporary" (e.g., Brindell, 1892; Adelaide's *Advertiser*, 12 September 1894; and subsequent reprints), "European and American astronomers" (Taylor, 1887; and subsequent reprints) or "European observers" (*The Age*, 24 September 1892; *Bathurst Free Press and Mining Journal*, 6 February 1893), or "the scientific minds of Europe and America" (Sydney *Daily Telegraph*, 25 August 1894), again without further bibliographic detail. In many cases, no attribution at all is provided, making it difficult for readers to distinguish between first-hand observation, second-hand reporting and editorial summary.

Within the present database, this pattern is especially evident in early Mars-related reporting from the late 1870s and early 1880s. Items describing the appearance of Mars, its satellites or surface markings are commonly presented as matters of fact, without explicit reference to individual observers. Authority is thus constructed institutionally or implicitly, rather than personally. The prevalence of anonymous or weakly attributed reporting reflects contemporary journalistic conventions rather than indifference to scientific credibility; on the routine anonymity and weak attribution of nineteenth-century newspaper reporting, see Aspinall (1949), Brake et al. (1990) and Secord (2004), who emphasise that authority was commonly conveyed through genre conventions and institutional reputation rather than explicit citation. In the absence of named individuals, authority is often conferred through indirect cues: references to "European astronomers", "leading observatories" or "recent scientific observations" proliferate. Such formulations signal reliability while avoiding the need for detailed sourcing. This pattern aligns with the treatment of astronomy as a largely consensual science during this period, in which observational results were presented as cumulative and non-contentious.

### 4.2 The emergence of named authority

Over time, however, Mars-related reporting exhibits a progressive shift towards explicit attribution. By the late 1880s and early 1890s, items increasingly name individual astronomers, particularly in connection with distinctive claims or discoveries. This shift is visible in reporting that references Giovanni Schiaparelli (1835–1910) in connection with Martian canals, where his name serves not only as a source but as a conceptual anchor for the discussion. Similarly, reports associated with organised observing programmes, such as those directed by Pickering at Arequipa, Peru, and Valparaiso, Chile, are more likely to identify the individual responsible, especially when observations are presented as systematic or coordinated across multiple sites. Named attribution in these cases functions as a marker of methodological seriousness and organisational scale.

By the mid-to-late 1890s, Mars reporting in the Australian press increasingly revolves around recognisable scientific personalities, most notably Lowell. Items discussing Martian canals during this period frequently emphasise Lowell's name in headlines or introductory sentences, sometimes presenting his views in direct or indirect quotation. This marks a significant departure from earlier reporting practices. Authority is no longer derived primarily from institutional consensus or anonymous expertise, but from the interpretive stance of an individual whose views are recognisable, repeatable and, at times, controversial. The prominence of Lowell's name allows newspapers to frame Mars-related material as part of an ongoing debate rather than as a sequence of observational updates. Importantly, such items often circulate widely even when they contain little new observational information. The authority of the named individual, instead of the novelty of the data, sustains their newsworthiness.

Explicit attribution also makes scientific disagreement more visible in the press, sometimes within the confines of a single article. A clear example appears in the interview titled *"The Planet Mars. Has It an Atmosphere. An Interesting Interview"* (*Australian Star*, 24 August 1894), which juxtaposes differing interpretations of recent Martian observations in direct succession. While reporting claims associated with American observers, the article immediately qualifies them by noting that "… in the colony … no one has seen anything of these markings" and adds that "… the question is being discussed amongst astronomers." Rather than presenting a settled conclusion, the piece explicitly acknowledges that "… there is no general agreement as to the result at present", allowing uncertainty and dissent to be articulated without destabilising the authority of astronomy as a field. The presence of multiple viewpoints within a single report marks a departure from earlier, more declarative Mars coverage.

By the late 1890s, Lowell's canal theories were increasingly reported alongside sceptical commentary, either from unnamed critics or from other astronomers identified by name. In several Lowell-related summaries, canals are described using distancing formulations such as "the so-called canals of Mars" or are followed by remarks that their interpretation is still questioned by many observers. Elsewhere, scepticism is attributed more explicitly. In the *Australian Star* interview, for example, Pickering is named as questioning prevailing interpretations and suggesting alternative explanations for the observed markings, while cautioning against premature conclusions. Such approaches allowed newspapers to register disagreement—whether anonymous or attributed—while preserving the collective credibility of astronomical inquiry. Dissent could thus be framed not as a challenge to science itself but as evidence of an active and ongoing process of expert evaluation. Astronomy is thus represented not just as a body of established facts but as an active field of inquiry marked by debate. This development further reinforces the role of attribution as a structuring device in science journalism.

The evolution of attribution practices has direct implications for the circulation patterns discussed in Section 3. Items anchored to anonymous or institutional authority tend to circulate as interchangeable informational units, whereas those associated with named individuals acquire a degree of narrative continuity. Readers encounter not just isolated reports but recurring figures whose views develop over time. From a methodological perspective, this shift underscores the importance of tracking names, as well as topics, within newspaper article collections. The emergence of personality-driven authority helps explain why certain Mars-related items persist in circulation long after their initial publication and why interpretive claims achieve prominence even in the absence of new observational triggers.

## 5. HALL, PICKERING AND LOWELL AS PRESS AUTHORITIES

### 5.1 Authority through discovery: Asaph Hall

The discovery of Mars's moons by Hall in 1877 was widely communicated in scientific circles almost immediately following the announcement from Washington, D.C., although the shape and longevity of his presence in the Australian press differed markedly from the unfolding local interest in Mars more generally. Hall's role in discovering Phobos and Deimos was reported less as part of an ongoing scientific odyssey and more as a discrete confirmation of a long-anticipated astronomical event—a completed achievement rather than a continuing commentary on Martian science. In Australia, as elsewhere, newspapers drew on overseas dispatches—largely from Europe and the USA—to inform readers of Hall's success in identifying the tiny satellites, e.g.:

> The great telescope of the [U.S.] Naval Observatory has just signalised itself, in the hands of Professor Hall, by one of the most remarkable additions to modern astronomy nothing less, in fact, than the discovery of one satellite, and probably two, to the planet Mars. (Henry, 1877).

These early notices typically explicitly referred to Hall and the discovery—often with succinct specifics like the dates (12 August 1877 for Deimos, 18 August 18 for Phobos) and the instrument used (the 26-inch/66 cm refractor at the U.S. Naval Observatory; e.g., *Sydney Mail and New South Wales Advertiser*, 27 October 1877)—but rarely followed up with detailed scientific analysis or further editorial engagement.

In the Australian colonial press of the late nineteenth century, Hall's name appears most frequently in brief notices or digests of international astronomical news. These often focused on the fact of the moons' existence rather than on Hall himself as a scientific personality. This reflects a broader reporting practice of the era, which we have encountered already, where colonial media transmitted significant scientific milestones but did not routinely engage in extended coverage of foreign scientists' careers or interpretative debates stemming from their work. For example, Australian papers ran items summarising the discovery amidst other science news—typically within the broader context of astronomical "firsts" and curiosities—rather than running sustained profiles or interviewing local experts in response to Hall's findings.

The pattern of reporting is characteristic: Hall's discovery is treated as a reportable fact, often appearing alongside short pieces on other celestial phenomena (such as Schiaparelli's observations of Mars's surface features), but with little continuation once the novelty wore off. This points to a mode of authority in which Hall's credibility was essentially locked into the successful "achievement" of identification—the immediate scientific milestone—without his later interpretations or ongoing scientific voice becoming matters of sustained public discussion in Australia.

Later press references to Hall, especially as Mars continued to occupy the public imagination, tended to invoke his discovery historically. That is, rather than quoting Hall's later scientific opinions or using him as a source in debates about Mars's habitability or geological features, Australian news and commentary invoked his name retrospectively to anchor discussions about Mars's observed features or the history of observations, e.g.:

> The last time an exceptionally close opposition occurred (in September, 1877), the renowned American astronomer, Asaph Hall, signalised the event by the discovery of two hitherto unknown satellites. (*Australian Star*, 26 August 1892).

In this way, Hall's authority in the press functioned less as a living voice and more as an event-based reference point.

By the early twentieth century, such retrospective invocation was common in Australian discussions of Mars. When debates about Mars's features intensified with increasing scientific scrutiny and public fascination with life on Mars (influenced by individuals like Lowell in the USA), Hall's name was used to legitimise the historical lineage of Martian observation rather than to participate in those debates itself. His role was cited as foundational rather than interpretive or theoretical. This pattern is significant because it illustrates how, within the Australian public sphere, scientific authority can be tethered to discrete achievements rather than sustained engagement with subsequent developments.

This reporting pattern reflects broader dynamics in nineteenth-century colonial journalism, where international scientific figures were often mediated through short, fact-focused reports rather than ongoing intellectual engagement. Hall did not become a household name in Australia in the way some contemporaries—such as Schiaparelli, with his vivid reports of *canali*—did. His one-off achievement was respected and acknowledged but not developed into a local touchstone for deeper Martian debate. Thus, Hall's presence in Australian newspapers demonstrates how event-based validation shaped scientific authority in the colonial press: his credibility was derived from having achieved something definite and reportable—the discovery of Mars's moons—and once that was widely accepted, there was little incentive for further defence or elaboration in the public domain. Only when later Martian phenomena captured Australian readers' imaginations did Hall's name re-emerge, historically anchored to the origins of Mars observation rather than as an ongoing scientific voice.

### 5.2 Organised observation and distributed authority: William H. Pickering

In contrast to Hall's episodic appearance in Australian newspapers, Pickering emerged in colonial news coverage as part of a broader and sustained narrative about the scientific observation of Mars itself. Pickering's footprint in the press was shaped less by singular revelations and more by ongoing observational activity, his methodological choices and his participation in organised campaigns that exploited the southern hemisphere's advantageous viewing conditions.

The 1892 opposition of Mars was widely anticipated within the global astronomical community, and it drew heightened media attention in Australia. Pickering's role was emphasised precisely because he had established himself at a southern observing station (Pickering, 1892), giving his work both geographical and practical relevance to southern-hemisphere audiences. In 1891 Pickering was sent by Harvard College Observatory to establish what became the Boyden Station at Arequipa, Peru, explicitly in advance of favourable global campaigns to observe Mars—the southern latitude offering clearer and higher-altitude views during opposition. This initiative itself attracted press coverage not simply as a scientific sidebar but as a tactical deployment of astronomical resources, highlighting location and infrastructure in the reporting rather than just focusing on the individual observer (see Paper 1).

Australian newspapers reporting on the 1892 opposition drew on dispatches circulated internationally which mentioned Pickering's efforts, "… who is well known for his work at Arequipa, in Peru, …" (*Leader*, 29 September 1894), to send regular telegrams back to readers about Martian surface features. These were not gathered as casual notes but as part of a coordinated observational campaign spanning continents, telescopes, and—importantly for the colonial press—competing geographical vantage points. The repeated reporting of his telegraphed observations to outlets such as the *New York Herald* (as cited by, e.g., Adelaide's *Evening Journal*, 15 October 1892; *Australische Zeitung*, 2 November 1892; and subsequent syndication) added a narrative of real-time engagement that resonated in the Sydney and Melbourne press cycles, keeping his name in the news across weeks (and not just around a single discovery moment).

Unlike Hall, whose authority in the Australian press was rooted in the successful completion of a particular discovery, Pickering was repeatedly invoked as part of the machinery of modern empirical astronomy. Terms that appear in colonial news linked Pickering's name with an assembled network of tools and observers, not just his personal expertise. The infrastructure of observation (telescopes, scheduled campaigns, southern positioning and data transmission) became a vehicle for establishing ongoing relevance in newspaper items. Through repeated cycles of reporting, his authority was constructed as both reliable and systematically engaged ("Professor Pickering has confirmed …"). This context was valuable in shaping Australian expectations: rather than presenting Pickering's work as isolated claims, newspapers framed his ongoing efforts as part of an international enterprise to map Mars's features across decades of opposition cycles. This reinforced a sense that Mars observation was an evolving project and that Pickering was a figure whose status was renewed at each relevant astronomical event, e.g.:

> It is rather amusing to hear of men like Mr. [Francis] Galton and Mr. [Hugh Reginald] Haweis[1] writing in the English papers about methods of signalling to the planet Mars merely because Professor Pickering has made out a little more detail in the vague image that the best telescopes give of the planet's surface. (*South Australian Register*, 5 January 1893).

> During the year 1892, Mars, one of the lesser planets of our own solar system, occupied a favorable [*sic*] position in the heavens for terrestrial observation. Owing to semi-scientific sensations concerning the supposed discovery of traces of civilisation on the surface of our celestial relation, the latter has become a subject of more than ordinary interest. This interest, owing to the theories put forward by Professor Pickering and others, has not been confined entirely to students of astronomy, but has been shared, to some extent, by the reading public. (*Illustrated Sydney News*, 18 February 1893).

While Hall's authority in the Australian press faded into historical reference after initial reporting, Pickering's reputation was continually refreshed. Even after the 1892 campaign ended—including when he was recalled from Peru and subsequently engaged in follow-up work (such as his contributions to later Martian surface studies and the burgeoning debate about canals and potential vegetation)—his name continued to appear in discussions in which Australian papers summarised northern- and southern-hemisphere perspectives on Martian features. Pickering's interpretations, including his acceptance of features like "lakes" and his willingness to discuss the potential implications of observed phenomena, kept him in the orbit of speculative discussions about Mars in the press.

> "Professor Pickering, of Harvard, the American astronomer, asserts that he has discovered 40 small lakes in the planet Mars." That's nothing! Last time our representative was on a visit to this planet he dropped across six seas, fourteen mountains, and a slygrog shop besides the lakes aforesaid; but the excursion was not fraught with much pleasure as it followed immediately on top of a banquet, and was mixed with snakes, blue-devils and onions! (*Nepean Times*, 15 October 1892).

The cumulative effect of this pattern is that Pickering's name functioned as an anchor linking local geographic advantage (being in a southern position) with international scientific endeavour. His repeated presence in news items, especially during and after key oppositions such as that of 1892, served to reinforce his standing as a reliable observational authority—not only because he obtained observations but because the structure and continuity of his work fit compellingly into the way Australian newspapers narrated the unfolding story of Mars. Pickering became, in the public imagination, part of the ongoing project of Martian study rather than a figure fixed to a momentary discovery. Whereas Hall's press identity was anchored in retrospective celebration, Pickering's was anchored in methodological continuity and repeated observational participation—a distinction that the Australian press reflected through greater frequency*,* contextual tie-ins to campaigns and repeated invocation across observation cycles.

## 5.3 Interpretive authority and continuity: Percival Lowell

In reporting on Mars during the late nineteenth and early twentieth centuries, Lowell represented a distinct and qualitatively different mode of scientific authority in Australian newspapers. Where Hall's presence was tied to a discrete discovery and Pickering's to ongoing observational campaigns, Lowell was invoked as a theorist and storyteller of Mars itself, a figure whose influence extended beyond the empirical to the interpretive and speculative. Lowell first began to appear in Australian press items in the mid-1890s, precisely as his works on Martian canals and habitability were disseminated internationally.

Contemporary accounts, including Australian reprints or summaries of foreign dispatches, often framed Lowell not simply as an observer but as an interpreter of Martian phenomena whose authority derived from his advocacy of a coherent interpretive position on the nature of Mars, e.g., "Mr. Lowell takes a different view of the seas to Mr. Pickering, and regards them as really water, …" (*Leader*, 29 September 1894). This interpretive authority contrasted sharply with the predominantly descriptive or factual reporting on Hall and Pickering. Central to Lowell's presence in the Australian press was his theory of Martian canals—the idea that a network of linear features on Mars was not a natural geological pattern but an artificial irrigation system built by intelligent beings. This hypothesis was widely known because of Lowell's books and lectures, beginning with *Mars* (1895; for context, see e.g. *The Australasian*, 25 July and 15 August 1896; *Australian Star*, 27 November 1896) and continuing through *Mars and Its Canals* (1906) and *Mars as the Abode of Life* (1908). His maps and detailed sketches of these putative canals were reproduced or described in international news accounts, which Australian newspapers often syndicated or summarised for local readers.

Unlike Hall, Lowell appeared repeatedly over an extended period, both in direct reporting of his latest claims and in reflective pieces linking his ideas to broader questions of Martian life (e.g., *Argus*, 16 November 1895; *Maitland Daily Mercury*, 4 February 1896). This ongoing presence shaped him in the press as more than an empirical reporter; Lowell became a name attached to a narrative about Mars, its features and its possible inhabitants. His interpretive framing was explicitly given space in Australian reporting: newspapers summarised his views on atmospheric conditions, the morphology of the Martian surface and the possible biological implications of his observations, often giving prominence to the life question ("Is our neighbour inhabited?") as a hook for readers.

Such coverage often mixed scientific reporting with broader cultural fascination: the idea of Martian life and engineered canals resonated with Victorian and Edwardian audiences as both a scientific possibility and a subject of imaginative inquiry. Lowell's continuous output (through books, lectures abroad and serialised summaries republished in overseas and colonial newspapers) meant that his name became shorthand for any serious or speculative discussion about Mars's surface and its implications:

> Mr. Lowell, the American astronomer, who observed Mars at Flagstaff, Arizona, where the atmosphere is wonderfully clear, has become convinced that the planet is inhabited by intelligent beings. According to a lecture which he delivered recently at Boston the "oases" connecting the canals are fertile spots of an oval or circular shape, usually about 100 or 150 miles [160–240 km] in diameter and of a green so dark that it is quite distinct in the telescope at the distance of 40,000,000 miles [64 million km]. He ascribes the color [*sic*] to vegetation, and states that it changes with season like that of the foliage on the [E]arth. (*Bendigo Advertiser*, 6 July 1895).

This contrasts sharply with Hall's discrete discovery notices or Pickering's episodic observational dispatches. Lowell's prominence in the press also invited response and debate. Australian newspapers occasionally paired reports of his claims with mention of sceptics who disputed the canal hypothesis or noted scientific controversy. This pattern underscores how his authority was recognised in context rather than assumed as uncontested fact. Instead of closing discussion after a single announcement, press cycles would revisit Lowell's ideas as Mars regained visibility in new oppositions or as new observational data (or counterclaims) emerged.

## 5.4 Comparative visibility

The differing roles of Hall, Pickering and Lowell are reflected in their relative visibility within the article compilation. Hall's name appears infrequently and is tightly bound to a specific historical moment. Pickering appears more regularly, particularly during periods of intensified observation, but remains associated primarily with empirical reporting. Lowell, by contrast, becomes a recurring presence whose name alone is sufficient to signal the topic and tone of an article. This progression mirrors the broader shift identified in earlier sections: from anonymous or institutional authority, through named observational credibility and, finally, to personality-driven interpretive authority. The press treatment of these personalities illustrates how different kinds of scientific authority are accommodated within journalistic conventions.

The form of authority associated with each person also influences the longevity of their presence in the press. Hall-related items tend to have a short circulation life, reflecting their factual and self-contained nature. Pickering-related items persist longer, often resurfacing across multiple reporting cycles as observational conditions recur. Lowell-related items exhibit the greatest persistence. Even in the absence of new observational data, references to Lowell's views continue to circulate, sustained by debate and speculation rather than any empirical update. This pattern reinforces the argument that personality-driven authority plays a key role in extending the lifespan of scientific topics within the press.

The contrasting press trajectories of Hall, Pickering and Lowell demonstrate that scientific authority is not monolithic in newspaper reporting. Different forms of authority—discovery, organised observation and interpretation—are accommodated in distinct ways, each with characteristic patterns of attribution, repetition and persistence. Recognising these differences is essential for understanding how astronomical knowledge was presented to, and consumed by, the Australian public. It also highlights the importance of attending to named individuals within any press compilation, not simply as sources of information but as structuring elements in the circulation of scientific ideas.

The emergence of personality-driven authority, particularly in the case of Lowell, cannot be understood in isolation from the mechanisms of news transmission that brought overseas science to Australian readers. The next section therefore examines the role of telegraphy, syndication and international news services in shaping the pathways through which Mars-related reporting entered and spread within the Australian press.

## 6. TELEGRAPHY, SYNDICATION AND COLONIAL CIRCULATION

### 6.1 International telegraphy and the arrival of astronomical news

By the late nineteenth century, Australian newspapers were firmly integrated into global telegraphic networks (e.g., Moyal, 1984; Potter, 2012). Submarine cables linked Australia to Europe and North America via Asia, enabling relatively rapid transmission of scientific news alongside political and commercial intelligence. Mars-related astronomical reporting entered the Australian press primarily through these channels, often in condensed or summarised form. Telegraphic transmission shaped the character of such reporting. Items received by cable tend to be brief, fact-focused and stripped of technical detail, a format well suited to initial announcement but less conducive to extended explanation. As a result, early Australian reports of Mars-related discoveries often appear as compact notices, later supplemented by longer articles derived from foreign newspapers or journals.

Analysis of first appearances indicates that Mars-related news typically entered Australia through metropolitan newspapers with established access to international cable services. Newspapers in Sydney and Melbourne are disproportionately represented among the earliest Australian publications of Mars-related items, reflecting their central role as gateways for overseas scientific news. Once published in these metropolitan centres, items were frequently reprinted by regional and provincial newspapers, either verbatim or in abbreviated form. This pattern reinforces the hierarchical structure of colonial news circulation, in which metropolitan papers functioned as intermediaries between global information networks and local readerships. Importantly, metropolitan mediation also involved editorial selection. Not all overseas astronomical news was transmitted or reproduced. Mars-related items that were deemed novel, dramatic or culturally resonant were more likely to be circulated widely.

### 6.2 Syndication and reprint chains

Beyond direct telegraphic transmission, Mars-related reporting often spread through syndication and informal reprint networks. Newspapers routinely borrowed material from one another, sometimes acknowledging the source but often without explicit attribution. These reprint chains can be traced in the database by comparing identical or near-identical texts across different titles and dates. Such chains reveal that a single overseas report, once introduced into the Australian press, could propagate extensively without further reference to its original source. In some cases, later reprints attribute the item to an Australian metropolitan newspaper rather than to the foreign origin, effectively domesticating the content. This process complicates efforts to reconstruct lines of influence but underscores the importance of identifying first appearances as anchors within the circulation network.

The mechanics of telegraphy and reprinting introduce characteristic temporal distortions into the circulation of Mars-related news. Initial cable reports may appear with minimal delay but in highly compressed form, whereas fuller accounts arrive later through non-telegraphic channels. Conversely, interpretive material may circulate long after its original publication overseas, sustained by reprinting rather than by new transmission. As items move through this system, they are subject to transformation. Technical detail may be omitted, interpretive passages emphasised and speculative elements made visible. These changes frequently reflect editorial judgements about audience interest and news value. The result is a layered press record in which different versions of the "same" Mars story coexist, each shaped by the constraints of the medium through which it travelled.

Although Australian newspapers relied heavily on overseas sources for Mars-related reporting, they should not be regarded as passive recipients. Editorial decisions regarding placement, headline framing and selection played a significant role in shaping how Mars was presented to readers. In some instances, Australian newspapers appended local commentary or contextual remarks to overseas material, particularly when discussing observational opportunities afforded by southern-hemisphere observers. Such interventions suggest an awareness of Australia's position within global astronomical networks and a focus on asserting local relevance. This mediating role reinforces the argument that colonial newspapers participated actively in the construction of astronomical knowledge rather than merely transmitting it unchanged.

The circulation mechanisms described in this section interact closely with the patterns of authority discussed in Sections 4 and 5. Anonymous or institutionally attributed items are particularly well-suited to telegraphic transmission and rapid reprinting, whereas personality-focused material benefits from longer formats and sustained circulation. Telegraphy thus facilitated the initial dissemination of Mars-related news, whereas syndication and reprinting enabled its persistence and transformation. Together, these mechanisms help explain the shifting prominence of different kinds of authority within the Australian press. The case of Mars illustrates how global scientific knowledge was routed through colonial press systems that combined technological connectivity with editorial agency. Telegraphic networks determined what could arrive quickly; reprint practices determined what endured. Understanding these circulation structures is essential for interpreting patterns of excitement, debate and authority in Australian Mars reporting. It also provides a framework for analysing other forms of scientific news within the colonial press.

## 7. TEMPORAL RHYTHMS OF MARS REPORTING

### 7.1 Episodic clustering and astronomical triggers

The temporal distribution of Mars-related items in the Australian press is characterised by distinct clusters of activity rather than continuous coverage. These clusters correspond most clearly to favourable oppositions of Mars, particularly those of 1877, 1888, 1892 and 1894, which act as external astronomical triggers for renewed reporting. In the immediate vicinity of such events, first appearances increase sharply, followed by periods of intense reprinting or syndication. This pattern reflects the responsiveness of the press to observational opportunity: Mars becomes newsworthy when its visibility improves, prompting both fresh reports and renewed circulation of existing material. The close alignment between oppositions and reporting peaks supports the conclusion that astronomical events, and not purely cultural fascination, initially structured the rhythm of Mars coverage.

Within each cluster, two temporal scales can be distinguished. The first is a short-term burst, typically spanning days or weeks, when newly arriving items circulate rapidly through metropolitan and regional newspapers. This phase is dominated by first appearances and near-simultaneous reprints. The second is a longer tail of persistence, when Mars-related material continues to appear intermittently long after the immediate observational context has passed. This long tail is especially pronounced for items involving interpretation, speculation or broader explanatory themes, such as discussions of Martian canals or habitability. The coexistence of these two temporal modes—burst and persistence—helps explain why Mars remained a visible topic in the Australian press even during periods when no new observational data were forthcoming.

In addition to astronomical triggers, Mars reporting exhibits seasonal and editorial rhythms associated with newspaper production cycles. Longer explanatory articles and essays are more likely to appear during periods of reduced political or commercial news pressure (although a full analysis of these aspects is beyond the scope of this paper), whereas brief notices cluster around moments of heightened international communication. Such rhythms are not unique to Mars reporting but interact with astronomical events to shape the overall temporal profile of the article database. The press thus imposes its own periodicity on scientific news, overlaying editorial cycles onto celestial ones (e.g., Secord, 2004).

## 7.2 Shifting temporal structure in the 1890s

A notable change occurs in the mid-to-late 1890s, when Mars-related reporting becomes less tightly coupled to specific oppositions. Although observational events continue to generate bursts of activity, a growing proportion of items circulate independently of immediate astronomical triggers. This shift is closely associated with the rise of personality-driven authority, particularly in relation to Lowell's interpretive claims. Articles discussing Mars increasingly appear as part of ongoing debates or thematic series rather than as responses to discrete events. As a result, the temporal structure of Mars reporting becomes more continuous and less episodic. This development marks an important transition in the public presentation of planetary astronomy.

Although a full comparative analysis is beyond the scope of this paper, the temporal behaviour of Mars-related reporting differs noticeably from that of many other astronomical subjects in the Australian press. Transient events such as eclipses or comets tend to generate sharp but short-lived peaks of attention, whereas Mars sustains interest across multiple years and observational cycles. This durability reflects the combination of recurring observational opportunities, speculative interpretive frameworks and the sustained circulation mechanisms described in earlier sections. Mars thus occupies a distinctive temporal niche within colonial science journalism.

The temporal rhythms identified here underscore the importance of analysing time as a structured dimension of newspaper reporting rather than as a neutral backdrop. Peaks, gaps and persistence patterns carry historical meaning and reflect the interaction of scientific events, media infrastructure and editorial judgement. By tracing these rhythms more directly, this study provides a framework for understanding how astronomical topics gain, lose and regain prominence in the public sphere. Such temporal patterns draw together the mechanisms of circulation, attribution and authority examined in earlier sections. In the next section, these findings are synthesised to consider what the case of Mars reveals more broadly about colonial science journalism and the public life of astronomy in late nineteenth-century Australia.

## 8. MARS AND COLONIAL SCIENCE JOURNALISM

### 8.1 Mars as a case study in press-mediated astronomy

It has become clear that Mars occupied a distinctive position within the Australian press in the last quarter of the nineteenth century. Unlike many astronomical topics that appeared only episodically, Mars generated repeated cycles of attention sustained by a combination of observational opportunity, interpretive flexibility and efficient mechanisms of circulation. As a case study, Mars reveals how astronomical knowledge was not just transmitted to colonial audiences but was actively shaped by the structures and practices of journalism. Telegraphy, syndication, attribution and editorial framing all played roles in determining which aspects of planetary astronomy became visible, persistent and meaningful in print.

The Australian press operated at the intersection of global scientific networks and local readerships. Overseas observatories and astronomers provided the raw material for reporting, but colonial newspapers selected, condensed and reframed that material according to local conventions and interests. This process had direct consequences for the construction of scientific authority. Anonymous or institutionally attributed reports dominated early Mars coverage, reflecting a model of authority grounded in collective expertise. Over time, however, overseas named individuals—most notably Pickering and Lowell—came to occupy increasingly prominent positions in press narratives. The shift from institutional to personality-driven authority did not replace earlier forms but layered upon them, producing a hybrid model in which discovery, observation and interpretation coexisted within the press record.

### 8.2 Temporal structure and public engagement

The temporal rhythms identified in Section 7 further illustrate how public engagement with astronomy was structured. Bursts of reporting around oppositions created moments of intensified attention, whereas often long tails of reprinting and commentary sustained interest across years. Such rhythms are not incidental but reflect the interaction of celestial cycles with journalistic practice. Recognising this interaction helps explain why certain astronomical topics achieved lasting prominence while others remained transient. From the perspective of the history of astronomy, this study underscores the importance of considering press mediation as part of the observational enterprise itself. Public understanding of Mars was shaped not only by what astronomers observed but by how those observations were communicated and debated in print. The Australian case highlights the global reach of nineteenth-century astronomy and the role of colonial contexts in amplifying, filtering and reframing metropolitan science. Mars-related reporting thus offers insights into the broader dynamics through which astronomical knowledge circulated beyond the observatory.

Whereas this paper has focused on the structural mechanisms underlying Mars reporting, the interpretive consequences of these mechanisms are explored in Paper 1. Together, the two studies demonstrate how circulation, authority and temporal structure conditioned the reception and meaning of Mars-related astronomy in Australia. The separation of methodological and interpretive analysis allows each paper to address a distinct set of questions while drawing on a shared evidentiary base. The findings presented here are necessarily shaped by the constraints of digitised newspaper sources and keyword-based retrieval. While the compilation captures a wide and representative range of Mars-related reporting, it cannot encompass every reference or nuance. Future work might extend this approach beyond 1899 to examine how emerging technologies, such as astrophotography, altered press mediation.

## 9. CONCLUDING THOUGHTS

This paper has examined the circulation of Mars-related astronomy in the Australian press between 1875 and 1899, with particular attention paid to the mechanisms that shaped its visibility, persistence and authority. Although this study focuses on Australia, the mechanisms identified here—telegraphic compression, syndication, attributional drift and personality-driven persistence—were characteristic of late nineteenth-century press systems across the broader British world and beyond.

By distinguishing first appearances from reprinting, tracing attribution practices and analysing temporal rhythms, I have shown that Mars reporting was structured by a complex interaction of astronomical events, media infrastructure and editorial judgement. Mars-related news typically entered Australia through metropolitan newspapers with access to international telegraphic networks, before spreading through syndication and informal reprint chains to regional and provincial titles. These processes amplified certain kinds of material—especially interpretive and speculative content—while allowing others to fade quickly. The press thus played an active role in shaping public encounters with planetary astronomy.

The prominence of attribution practices across my article collection also suggests that readers were expected to recognise and differentiate between scientific authorities, indicating a level of familiarity with astronomical debate that shaped how credibility was assessed. Over time, patterns of attribution evolved from anonymous or institutionally grounded authority towards a more personality-driven model. The contrasting press trajectories of Hall, Pickering and Lowell illustrate how different forms of scientific authority—discovery, organised observation and interpretation—were accommodated within journalistic conventions, each with characteristic patterns of repetition and longevity. Although Australian newspapers drew extensively on international sources, their treatment of figures such as Lowell, Pickering and Flammarion demonstrates that colonial readers encountered a heterogeneous field of authority rather than a single dominant voice. The temporal structure of Mars reporting further underscores the importance of press mediation. Whereas astronomical oppositions triggered bursts of coverage, long tails of reprinting and commentary sustained interest across years, particularly in the late 1890s. Mars thus became a durable object of public scientific attention, not solely because of new observations but because of how those observations were circulated and reframed.

Taken together, these findings demonstrate that the public life of nineteenth-century astronomy cannot be understood without close attention to the press systems through which it moved. The Australian case highlights the role of colonial newspapers as both conduits and mediators of global scientific knowledge. By focusing on circulation rather than interpretation, this paper lays the methodological groundwork for the companion study (Paper 1), which explores how these mechanisms shaped the meanings attached to Mars within Australian society.

## 10. NOTE

1. The Rev. Hugh Reginald Haweis (1838–1901), an English Anglican clergyman, writer and public lecturer, gained attention in the 1890s for his speculative proposals on interplanetary communication, including the idea of signalling to Mars using powerful lights such as coordinated streetlamps in London. Sir Francis Galton (1822–1911), a British polymath and cousin of Charles Darwin, is known for his work in statistics, geography and public science discourse. In the context of Pickering's observations, both Haweis and Galton were cited in Australian newspapers as commentators on the possibilities of communicating with Martians, reflecting broader Victorian interest in the implications of Mars observation beyond empirical astronomy. Their appearances alongside Pickering illustrate how Australian press coverage of the 1892 opposition combined rigorous observation with speculative discussion on life and communication on Mars.